\documentclass[prl,a4paper,twocolumn,showpacs,superscriptaddress,floatfix]{revtex4}
\usepackage{graphicx}
\newcommand{\be}{\begin{equation}}
\newcommand{\ee}{\end{equation}}

\begin{document}

\title{Fluctuation Phenomena in Chaotic Dirac Quantum Dots: Artificial Atoms on Graphene Flakes}

\author{J. G. G. S. Ramos}
\affiliation{Departamento de F\'isica, Universidade Federal da Para\'{\i}ba, 58051-970, Jo\~ao Pessoa, Para\'iba, Brazil}
\author{M. S. Hussein}
\affiliation{Instituto de Estudos Avan\c cados and Instituto de F\'{\i}sica, Universidade de S\~{a}o Paulo, C.P.\ 66318, 05314-970 S\~{a}o Paulo, SP, Brazil.\\
 Departamento de F\'{i}sica, Instituto Tecnol\'{o}gico de Aeron\'{a}utica, CTA, S\~{a}o Jos\'{e} dos Campos, S.P., Brazil.}
\author{A. L. R. Barbosa}
\affiliation{Departamento de F\'isica, Universidade Federal Rural de Pernambuco, Dois Irm\~aos, 52171-900 Recife, Pernambuco, Brazil}

\date{\today}

\begin{abstract}

We develop the stub model for the Dirac Quantum Dot, an electron confining device on a grapheme surface. Analytical results for the average conductance and the correlation functions are obtained and found in agreement with those found previously using semiclassical calculation. Comparison with available data are presented. The results reported here demonstrate the applicability of Random Matrix Theory in the case of Dirac electrons confined in a stadium.

\end{abstract}
\maketitle

\section{Introduction}
The electronic transport across a wide class of recently controlled materials displays relativistic properties, despite its dynamics presents a speed much lower than the light. These structures are known as Dirac materials \cite{Firsov,Kim,Novoselov,Geim,beenakkergrafeno,Ensslin,Stampfer,Balatsky,Hankiewicz} and give rise to intriguing physical phenomena of interest both experimental and theoretical \cite{Geim1,Savchenko, Folk,Lerner,Roche,Xiong}. Quite interesting phenomena emerge from the nature of wave functions of the confined electrons, described by massless or massive Dirac equation of relativistic quantum mechanics \cite{beenakkergrafeno,Hankiewicz,Altshuler,Guinea,Baranger,Sigrist,Vladimir,Ostrovsky,Adagideli}, instead of the Schr\"{o}dinger equation. 

The Dirac equation is appropriate to describe the electronic states of two independent sub-lattice components \cite{beenakkergrafeno,JacquodButtiker}, which generates additional constraints known as pseudo-spins. The prominent examples of these bipartite systems are square lattices, such as some topological insulators, and hexagonal lattices, whose main example are the graphene structures.

Among the different electronic Dirac devices, the chaotic Dirac quantum dot (DQD), also called chaotic Dirac billiard (DB), has received a significant highlight\cite{Geim,Baranger,Adagideli1,Grebogi,Rycerz,Barros,Brouwer,Stampfer1,Guinea1,Guo,Subramaniam,Efetov}, due to its universal characteristics. In the search for such universal properties, the Ref.~[\onlinecite{Geim}] studies experimentally a DB using a graphene quantum dot. The authors study a small billiard and show level statistics distribution best described by Gaussian unitary or orthogonal ensembles from the Random Matrix Theory (RMT). Moreover, the authors find evidences of a time reversal symmetry (TRS) broken state in the absence of a magnetic field, raising questions about the possible origin of such states. In fact, almost thirty years ago, Berry and Mondragon \cite{Berry87}, studied what they called ``Neutrino Billiard", a stadium where massless spin-1/2 fermions, described by a Dirac Hamiltonian, are  confined. They showed that the system exhibits time reversal symmetry (TRS) breaking in the absence of an external magnetic field. The statistics of the energy eigenvalues  of the confined fermions were found to be governed by the Gaussian Unitary Ensemble (GUE). Quite recently, the Dirac stadium was experimentally studied with the aid of microwave resonators in Refs. \cite{Richter1, Richter2}. Motivated by  \cite{Berry87} and other findings, Ref.[\onlinecite{Baranger}] uses the tight-binding Dirac Hamiltonian model for electrons, taking into account massive confinement, to analyze the conductance and energy level statistics of graphene DB. In the absence of massive confinement, the authors show that electronic properties are well described by Gaussian unitary or orthogonal ensembles, as obtained in Ref.[\onlinecite{Geim}]. However, in the presence of massive confinement, the transmission statistics follow exclusively from the block unitary structure, while the spectral statistics exhibits an orthogonal or even Poisson statistics.

Analytical results for the chaotic graphene quantum dot (with massive confinement) were obtained using semiclassical theory in Ref.[\onlinecite{Adagideli1}]. In the limit of high massive confinement, the authors predict the average of conductance and the amplitude of the universal fluctuations as a function of magnetic flux and armchair edges. The authors also analyze how the ratio between the dwell time and the magnetic flux time ($T_{dwell}/T_{{\cal B}}$) as well as the ratio of the dwell time and armchair edges time ($T_{dwell}/T_{ac}$) affect the weak localization and universal fluctuations in the crossover regime and compare with standard results of the universal Gaussian Unitary and Orthogonal ensembles. Motivated by that semi-classical theory analysis, Ref.[\onlinecite{Barros}] performed a full analytical study of the DQD through the RMT chiral ensemble. The authors derived a general expression for the average conductance and its universal fluctuations for the three classes of chiral ensembles in the pure regime (in the absence of crossover fields), which in the semi-classical limit (large number of open channels) recover the specific results of Ref.[\onlinecite{Adagideli1}] that are outside the crossover regime.

As previously discussed, there is a theoretical challenge to construct an RMT formulation for the study of the DQD in any crossover regime (finite field and/or boundary condition). In order to solve the problem, we deduce in this work a generalization of the crossover scattering framework, based on a diagrammatic method which was proposed in Ref.[\onlinecite{Halperin}], to study the crossover regime in the chaotic Dirac billiard connected to a source and a drain of the conductance electrons. As a relevant application of our framework, we study the chaotic graphene quantum dot (chaotic DQD), obtaining  general analytical expressions for the average of the conductance and for the correlation functions of conductance as a function of energy, magnetic flux, straining of the graphene monolayer\cite{Geim1,Guinea} and confinement by massive, armchair and zigzag edges \cite{beenakkergrafeno,Richter}. In particular, we show the full suppression of the weak localization term as a function of the straining, in full agreement with the experimental findings of Ref.[\onlinecite{Geim1}]. Moreover, in the limit of high massive confinement, we recover the results obtained in Ref.[\onlinecite{Adagideli1}] which uses semi-classical theory. However, we emphasize the generality of our RMT framework, which is applicable to all categories of chaotic DQD.

The work is divided as follows: In Section I we give a brief account of the Quantum Chaotic Scattering Theory employed in the RMT treatment of chaotic Dirac quantum dot (CDQD). In Section II, we introduce the RMT crossover scattering framework to the chaotic Dirac billiard. In Section III, we include a brief discussion about the effective graphene Hamiltonian and apply the it to discuss the statistical properties of the CDQD. We perform calculations and obtain general analytical expressions for the average of conductance and its correlation functions and also analyze their relevant limits. The conclusions are given in Section IV.

\section{Quantum Chaotic Scattering Theory and the RMT-Based Stub Model}\label{Scattering}

In a general stadium which confines electrons one can describe the conductance and its universal fluctuations using known methods of resonant scattering. The electrons inside the stadium execute confining potential-affected  motion governed by the Schr\"dingier or the Dirac equation. The electrons suffer multiple reflections from the boundaries and standing waves are generated, which represent the eigenstates of the system. Taking into account the coupling of the interior of the stadium  to the outside region results in transforming the standing waves into resonances with a width that measures the time it takes the electrons to be transmitted yo the outside and electric conductance ensues. A convenient way to formalize the above, is through Feshbach's projection operator method, commonly used to treat the compound nucleus resonances in nuclear reactions \cite{Fesh1, Fesh2, Fesh3, CH13}. Denoting the total wave function of the system by $|\Psi>$, one introduces the projector $Q$ which projects out the closed channels, namely the states in the interior of the stadium. The states with electrons outside the stadium, namely the open channels are projected out by $P$, with $PQ=QP=0$ and $P^2 =P$, and $Q^2=Q$. The wave equation of the whole many-electron system, can then be decomposed into two coupled equations, one for $P\left| \Psi \right>$, and the other for $Q \left| \Psi \right>$. The exact, full Hamiltonian of the system $H$, is also decomposed into four operators, viz $QHQ + PHP + PHQ + QHP$. After well known manipulations one is able to derive a general exact expression for the scattering matrix $S$, that describes resonant scattering,

\begin{equation}
S = 1 -2i\pi PHQ\frac{1}{E - QHQ - QHP\frac{1}{E - PHP + i\varepsilon}PHQ}QHP
\end{equation}
Writing $QHP[1/(E-PHP + i\varepsilon)]PHQ = -i\pi QHP \delta(E-PHP)PHQ + QHP Pr[1/(E-PHP)]PHQ \equiv -i\Gamma_{Q}/2+ \Delta_{Q}$, where $\Gamma_{Q}$ is the width operator of the resonances, and $\Delta_{Q}$ is the real energy shift operator which is usually added to $QHQ$ to define the resonance Hamiltonian. Thus,

\begin{equation}
S = 1 -2i\pi PHQ\frac{1}{E - [QHQ +\Delta_{Q}] + i\frac{\Gamma_{Q}}{2}}QHP
\end{equation}

The above expression of the $S$-matrix is exact. For application to a given physical system, one has to specify the Hamiltonian $QHQ$ os the isolated closed stadium or quantum dot., and use a spectral decomposition of $\delta(E -PHP)$. This is accomplished in \cite{Mahaux}, and used extensively by \cite{Weiden1}. Neglecting the energy shift operator, and using matrix notation, the $S$-matrix which constitutes the basic theoretical object in Quantum Chaotic Scattering Theory based on RMT is

\begin{equation}
S = 1 -2i\pi W^{T}\frac{1}{E - H + i\pi W \dot W^{T}}W, \label{S-H}
\end{equation}
where $W$ is a real non-random matrix that represents the coupling of the internal region with the open channels, and $H$ is taken as a random Hamiltonian pertaining to one of the university classes of random matrices, the Gaussian Orthogonal Ensemble (GOE) with TRI, the Gaussian Unitary Ensemble, with TRI breaking, and the Gaussian Simplectic Ensemble (GSE). 
When using one of these ensembles to calculate averages of $SS^{\dagger}$ or $SSS^{\dagger}S^{\dagger}$, the distribution $P(H)$ is required. Analytical evaluation of these averages is quite involved as they require the evaluation of complicated triple integrals. Only in the case of GUE was it possible to actually obtain closed form expressions \cite{Weiden1}. Generally, researchers rely on numerical simulations using random matrix generator \cite{Weiden2}. Application of this theory to to microwave resonator physics is an ongoing program \cite{Weiden3}.

An alternative method which allows the obtention of analytical results for any of the ensembles is based on the distribution of the $S$-matrix itself, $P(S)$ \cite{Mello}. The ensemble here is Dyson's  Circular Unitary Ensemble (CUE). The stub model is based on this approach, see \cite{stub1}, and it amounts to attaching a stub (fictitious) to the quantum dot and use it as a scaffolding to build the S-matrix. The size of the stub is chosen so that the dwell time in it is much larger than the dwell time in the dot. Further, the permanence time of the electrons inside the dot + stub system, $\tau_{p}$, is much shorter than the escape time, $\tau_{esc}$ to the leads and open channels. These conditions guarantee that all the variables in the system, the Fermi energy $\varepsilon$, and the magnetic field, $B$, are made to be explicitly present in the reflection matrix $R$ (of dimension (M - N)) of the stub, leaving the S-matrix of the dot (without the stub), $U$, of dimension $M \times M$, be a a product of $2X2$ spin matrix times a matrix at $\varepsilon = 0$, $B = 0$ and zero spin-orbit scattering rate. As such the dot $S$-matrix, $U$ can be chosen from Dyson's Circular Orthogonal Ensemble of random matrix theory. It has been shown \cite{stub1} that the $S$-matrix of the system (dot plus leads) is
\begin{equation}
S = PU(1 - Q^{\dagger}RQU)^{-1}P^{\dagger}
\end{equation}
where $P$ and $Q$ are projection matrices of dimensions, $N \times M$ and $(M - N) \times M$, respectively. It has been proven that owing to the second condition on the time scales, namely,  $\tau_{p} \ll \tau_{esc}$,  the $S$-matrix above remains unaffected by the stub and in fact equivalent to the Hamiltonian-based $S$-martix, Eq.(\ref{S-H}), \cite{stub2}. Thus the characterization of the stub as a scaffolding is appropriate. To perform averages of $S$, one expands in powers of $U$ and uses diagrammatic techniques as developed by \cite{stub3}.\\

We turn now to a Dirac version of the $S$-matrix distribution approach and the stub model. 

\section{Crossover Scattering Framework for the General Dirac Billiard}\label{Crossover}

In this section, to study the crossover regime in chaotic Dirac billiard connected to leads, we introduce a generalization of the crossover scattering framework, which was proposed in Ref.[\onlinecite{Halperin}]. We begin by employing Quantum Chaotic Scattering Theory, and introduce the stub model discussed in the previous section for the chaotic Schr\"dingier quantum dot. Within the stub model the scattering matrix as function of external parameters such as the energy $\epsilon$ and magnetic flux $\mathcal{B}$, and an an internal parameter which is  the massive mass term, $m$, in the generic Dirac Hamiltonian, is given by
\begin{eqnarray} \mathcal{S}(\epsilon,\mathcal{B},m)=\mathcal{P}\left[\mbox{$\openone$} -\mathcal{Q}^{\dagger}\mathcal{R}(\epsilon,\mathcal{B},m)\mathcal{Q}\mathcal{U}\right]^{-1}\mathcal{U}\mathcal{P}^\dagger. \label{S}
\end{eqnarray}
The matrices $\mathcal{S}(\epsilon,\mathcal{B},m)$ and $\mathcal{U}$ have dimension $N_T\times N_T$ and $M\times M$, respectively. The total number of open channels $N_T=N_1+N_2$ is the sum of open channels in the leads $1$ and $2$, while $M$ is the number of resonances in the chaotic Dirac quantum dot. The matrix $\mathcal{U}$ can be a member of the Circular Orthogonal Ensemble instead of Chiral Orthogonal Ensemble if we assume $N_T\gg 1$. In this limit, Chiral universality classes give the same results of Wigner-Dyson universality classes as proved in Ref. [\onlinecite{Barros}]. The matrices $\mathcal{P}$ and $\mathcal{Q}$ are the projector operators of order $N_T \times M$ and $(M-N_T)\times M$, respectively, with elements given by $\mathcal{Q}_{ij}=\delta_{i+N_T,j}$ and $\mathcal{P}_{ij}=\delta_{ij}$. We intend to incorporate additional degrees of freedom on the formalism, two for each subspace, with the prominent example the structure of the graphene Hamiltonian. Accordi
 ngly, the elements of the matrices $\mathcal{U}$, $\mathcal{P}$ and $\mathcal{Q}$ are all proportional to the $\sigma_{0} \otimes \tau_{0}$, with $\sigma_{0}$ and $\tau_{0}$ denoting $2\times 2$ identity matrices. The matrix $\mathcal{R}(\epsilon,\mathcal{B},m)$ has dimension $(M-N_T)\times (M-N_T)$ and is parameterized as
\begin{equation}
\mathcal{R}(\epsilon,\mathcal{B},m)=\exp\left\{\frac{i}{M}\left[2\pi\frac{\epsilon}{\Delta}\sigma_{0} \otimes \tau_{0} -\mathcal{H}(\mathcal{B},m)\right]\right\}.\label{RH}
\end{equation}
In Eq.(\ref{RH}), the universal hamiltonian $\mathcal{H}(\mathcal{B},m)$ is obtained from the effective Dirac Hamiltonian preserving its intrinsic symmetries and considering its amplitudes as members of a Gaussian distribution. We consider the additional degrees of freedom residing in the elements of matrices $\mathcal{H}(\mathcal{B},m)$ which are all proportional to $\sigma_{i} \otimes \tau_{j}$, with $\sigma_{i}$ and $\tau_{j}$ denoting Pauli matrices ($i,j=x,y,z$) in each subspace of the Dirac Hamiltonian.
\begin{figure}[h]
\begin{center}
\includegraphics[width=8cm,height=4cm]{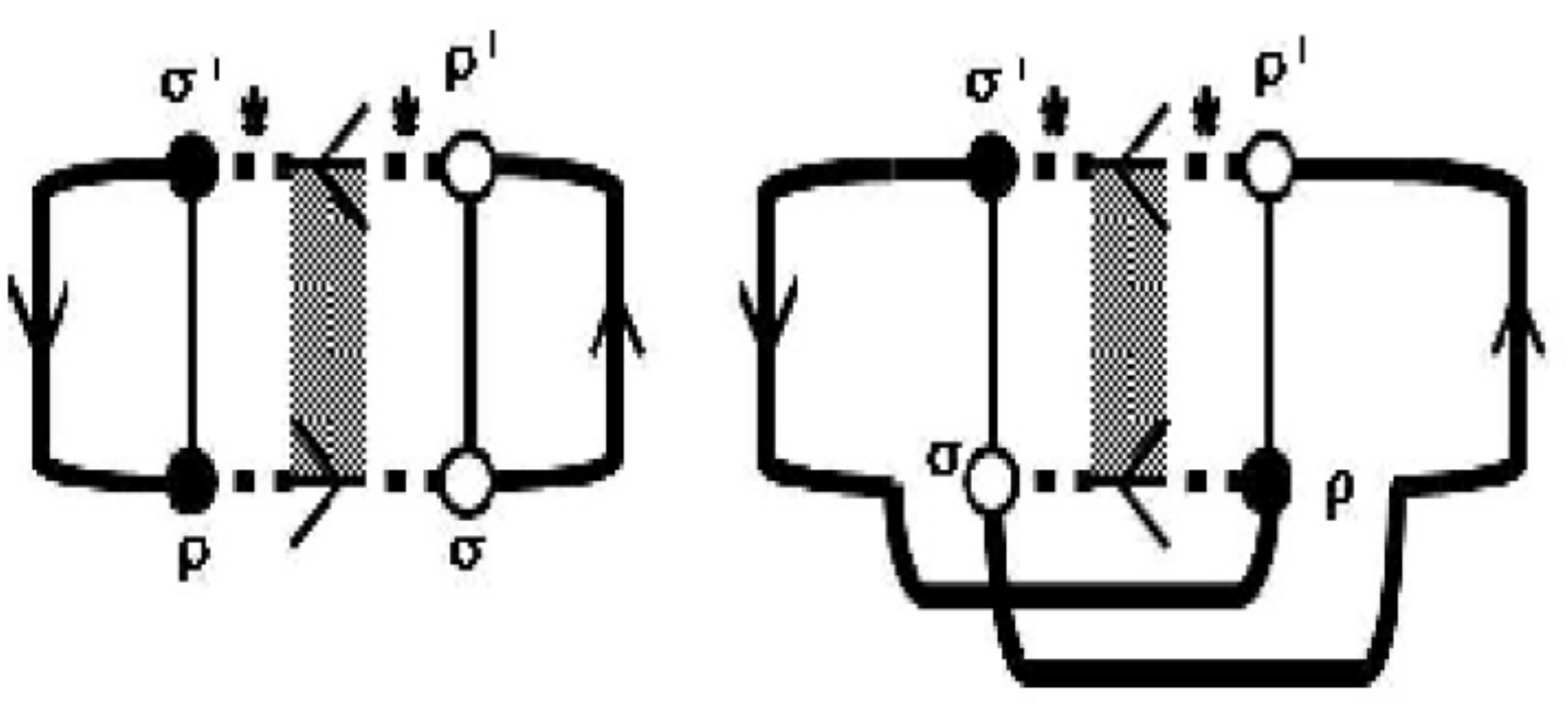}
\end{center}
\caption{Diffuson (left) and cooperons (right) diagrams for the average of conductance, Eq.(\ref{G}).} \label{diagramagwl}
\end{figure}

The conductance of the chaotic Dirac quantum dot at zero temperature can be written as a function of the scattering matrix, Eq. (\ref{S}), as follows
\begin{equation}
\frac{G}{e^2/h}=4 \times \frac{N_1N_2}{N_T}+ {\bf Tr}\left[ \mathcal{S}\;\mathcal{K}\; \mathcal{S}^\dagger\;\mathcal{K}\right].\label{G}
\end{equation}
where the elements of the matrix $\mathcal{K}$ are $\mathcal{K}_{ii}=N_2/N_T$, $\mathcal{K}_{ii}=-N_1/N_T$ and $\mathcal{K}_{ij}=0$ for $i=1,\dots,N_1$, $i=N_1+1,\dots,N_T$ and $i\neq j$, respectively. The factor $4$ arises from the degeneracies  of the two subspaces represented by $\sigma$ and $\tau$. The average of the conductance, Eq.(\ref{G}), can be obtained using the relation
\begin{eqnarray}
&& \left< \mathcal{S}_{ij;\alpha \beta;\rho \delta}(\epsilon,\mathcal{B},m) \mathcal{S}_{i'j';\alpha' \beta';\rho' \delta'}^*(\epsilon',\mathcal{B},m)\right> = \nonumber \\
&& \delta_{ii'}\delta_{jj'}\mathcal{D}_{\alpha \beta,\rho \delta;\beta' \alpha',\delta' \rho'} + \delta_{ij'}\delta_{ji'}({\cal T}\mathcal{C}{\cal T})_{\alpha \beta,\rho \delta;\beta' \alpha', \rho' \delta'}, \nonumber\\\label{SS}
\end{eqnarray}
whose validsty is over the limit $M\gg N_T\gg 1$. The ${\cal T}$ carries symmetries as the time-reversal of the Dirac Hamiltonian and is defined as ${\cal T}=\sigma_{0} \otimes \tau_{0} \otimes {\cal T}_{\chi}$. The indices $\alpha,\beta=1,2$ are associated with the subspace $\sigma$, while $\rho, \delta=1,2$ with the subspace $\tau$. The matrices $\mathcal{D}$ and $\mathcal{C}$ are contributions of diffuson and cooperon diagrams, which is represented in the Fig. (\ref{diagramagwl}), and obtained by following expressions

\begin{eqnarray}
    \mathcal{D}^{-1}&=&M \sigma_{0} \otimes \tau_{0} \otimes \sigma_{0} \otimes \tau_{0} -{\bf Tr} \left(\mathcal{R} \otimes \mathcal{R}^{\dagger} \right), \nonumber \\
    \mathcal{C}^{-1}&=&M \sigma_{0} \otimes \tau_{0} \otimes \sigma_{0} \otimes \tau_{0}-{\bf Tr} \left(\mathcal{R} \otimes \mathcal{R}^{\star}\right), \label{DC}
\end{eqnarray}
where $\dagger$ designates Hermitian conjugation and $\star$, complex conjugation. From Eqs. (\ref{G}) and (\ref{SS}), we can obtain the following expression for the average conductance:
\begin{equation}
\frac{\left<G\right>}{e^2/h}=4 \times \frac{N_1N_2}{N_T} -\frac{N_1N_2}{N_T} \sum_{\rho,\delta} \left[{\bf Tr}\left({\cal T}\mathcal{C}{\cal T}\right)\right]_{\rho \sigma; \rho\delta }, \label{Gm}
\end{equation}
where the trace involves the two subspace using the following general cross product
\begin{eqnarray}
\left[ {\bf Tr} \left( \sigma_{i} \otimes \tau_{j} \otimes \sigma_{k} \otimes \tau_{l} \right) \right]_{\rho \delta;\rho' \delta'} =\nonumber\\\left[\left(\sum_{\alpha \beta} (\sigma_{i})_{\alpha \beta} (\sigma_{k})_{\beta \alpha}\right) \tau_{j} \otimes \tau_{l} \right]_{\rho \delta,\rho' \delta'} '.\nonumber
\end{eqnarray}
The calculation algebra of Eq. (\ref{Gm}) is performed using the backward multiplication as following
\begin{eqnarray}
&&(\sigma_{i}\otimes\tau_{j}\otimes\sigma_{k}\otimes\tau_{l}) \cdot (\sigma_{i'}\otimes\tau_{j'}\otimes\sigma_{k'}\otimes\tau_{l'}) = \nonumber \\
&&(\sigma_{i} \sigma_{i'}) \otimes (\tau_{j'} \tau_{j}) \otimes (\sigma_{k'} \sigma_{k}) \otimes (\tau_{l} \tau_{l'}). \nonumber
\end{eqnarray}

The same algebraic analysis can be applied to the covariance of conductance. We perform the calculation and, after some algebra, we obtain the following expression
\begin{equation}
    \frac{{\textrm cov}\left[G(\epsilon, \mathcal{B}),G(\epsilon', \mathcal{B}'\right]}{e^4/h^2}=\frac{N_1^2N_2^2}{N_T^2}\left[{\cal V}_{{\cal D}}+{\cal V}_{{\cal C}}\right] ,\label{covg}
\end{equation}
where
\begin{eqnarray}
    {\cal V}_{{\cal D}}&=&\sum_{\rho, \sigma} \left[ {\bf Tr}{\cal D} \right]_{\rho \sigma; \rho' \sigma'} \left[ {\bf Tr}{\cal D} \right]_{\sigma' \rho'; \sigma \rho}, \nonumber \\
    {\cal V}_{{\cal C}}&=&\sum_{\rho, \sigma} \left[{\bf Tr}({\cal T}{\cal C}{\cal T}) \right]_{\rho \sigma; \rho' \sigma'} \left[ {\bf Tr}({\cal T}{\cal C}{\cal T}) \right]_{\rho' \sigma'; \rho \sigma}. \nonumber
\end{eqnarray}

To finalize this section, we can conclude that the crossover scattering model presented above is general and applicable to everyone kind of chaotic Dirac quantum dot. We need only getting the matrix ${\cal H}(\mathcal{B},m)$ from Dirac Hamiltonian together with Eqs. (\ref{RH}), (\ref{DC}), (\ref{Gm}) and (\ref{covg})  to obtain the averages of conductance and covariance. In the next section, we will apply the framework in the relevant example of a chaotic graphene quantum dot. We will present general results and, at specific limits, we recover the results of Ref. [\onlinecite{Adagideli1}].

\section{Chaotic Graphene Quantum Dot}\label{graphene}

In this section, we apply the crossover scattering model, which was described in the previous section, to study a general chaotic graphene quantum dot. First, the effective Hamiltonian of graphene is presented together with the symmetries of the problem. Following this, the characteristic and general effective graphene matrix, $\mathcal{H}(\mathcal{B},m)$, is introduced  and used in the calcukation of the average of conductance and covariance, Eqs.(\ref{Gm}) and (\ref{covg}), respectively.

\subsection{Effective Hamiltonian of Graphene}

Following Refs.[\onlinecite{beenakkergrafeno,Richter}], the effective Hamiltonian of graphene for low energies and long lenght scales whithout spin degree freedom can be written as
\begin{eqnarray}
    \mathcal{H}_{eff}&=&v\left[ {\bf p}-e{\bf A} \cdot {\bf \sigma}\right] \otimes \tau_{0}+e v \left[{\bf{\it A}}({\bf r}) \cdot {\bf \sigma}\right] \otimes \tau_{z}\nonumber\\
&+& w_{ac}({\bf r})\sigma_{z} \otimes \tau_{y}+m({\bf r}) \sigma_{z} \otimes \tau_{z} \nonumber\\
&+& w_{zz}({\bf r}) \sigma_{z} \otimes \tau_{z} \label{H}
\end{eqnarray}
where the Pauli matrices $\sigma_i$ and $\tau_i$ act on the sub-lattice and valley degrees of freedom, respectively. The vector potential ${\bf A} $ carries information about the external electromagnetic fields, and has no role in coupling the two valleys. The two valleys are coupled by a valley-dependent vector potential $\bf{\it A}({\bf r})$ produced by straining the monolayer \cite{Geim1,Guinea}. The boundary of chaotic graphene quantum dot is described by three physically relevant boundary types, which known as confinement by the mass term ($m({\bf r})$) , confinement by the armchair edges term ($ w_{ac}({\bf r})$), confinement by the zigzag edges term. However, there are four anti-unitary symmetries operating in graphene: ${\cal T}_{\chi}=\sigma_{y} \otimes \tau_{\chi}C$ with $\chi=\{0,x,y,z\}$,  with $C$  the operator of complex conjugation. ${\cal T}_{y}$ is the time reversal operation that interchanges the valleys,  while ${\cal T}_{x}$ is the valley symmetry. ${\cal T}_{0}$  is called a symplectic  symmetry, does not interchange the valleys and  is broken by massive term and  valley-dependent vector potential.

\subsection{Average of Conductance}

The central feature responsible for the simplified random-matrix description of the crossover in the universal regime is the fact that all relevant time scales are much bigger than the electron transit time $T_{erg}$, thus $T_{\cal B},T_{st},T_{ac},T_{m},T_{zz}\gg T_{erg}$. The significance of the crossover effect is guaranteed by the requirement that $T$'s are of the order of the inverse mean level spacing in the chaotic graphene quantum dot (also called chaotic Dirac quantum dot, CDQD). We may thus introduce the following dimensionless parameters to characterize the intensity of symmetry breakings in the system:

$$x^{2} = \frac{2\pi\hbar}{\Delta T_{{\cal B}}},\;\;w_{st}^{2} = \frac{2\pi\hbar}{\Delta T_{st}},$$

$$w_{ac}^{2} = \frac{2\pi\hbar}{\Delta T_{ac}},\;\;m^{2} = \frac{2\pi\hbar}{\Delta T_{m}},\;\;w_{zz}^{2} = \frac{2\pi\hbar}{\Delta T_{zz}}$$
where $\Delta$ is the mean level spacing.
From Eq. (\ref{H}), the random-matrix models for the effective Hamiltonians of graphene then follow directly from general symmetry considerations.
They are given by
\begin{eqnarray}
   \mathcal{H}&=&ix\; A_{1}\;\sigma_{x} \otimes \tau_{0}+ix\; A_{2}\;\sigma_{y} \otimes \tau_{0} \nonumber \\
&+&iw_{st}\; B_{1}\;\sigma_{x} \otimes \tau_{z}+iw_{st}\; B_{2}\;\sigma_{y} \otimes \tau_{z}\nonumber\\
 &+&
iw_{ac}\; Y\;\sigma_{z} \otimes \tau_{y}+ im X\;\sigma_{z} \otimes \tau_{z}\nonumber\\
&+& iw_{zz}\; Z\;\sigma_{z} \otimes \tau_{z} \label{HB}
\end{eqnarray}
where the matrices $A_{i}$, $B_{i}$, $X$, $Y$ e $Z$ are real antisymmetric with $\left<{\bf Tr}(A_{i}A^{T}_{j})\right>=\left<{\bf Tr}(B_{i}B^{T}_{j})\right>=\delta_{ij}M^{2}$ e $\left<{\bf Tr}(XX^{T})\right>=\left<{\bf Tr}(YY^{T})\right>=\left<{\bf Tr}(ZZ^{T})\right>=M^{2}$.

Now, we can substitute Eq.(\ref{HB}) in Eqs.(\ref{RH}) and (\ref{DC}) and after some algebraic manipulations, we obtain,
\begin{eqnarray}
{\cal D}^{-1}={\cal C}^{-1} &=&{\cal N} \; \left(\sigma_{0} \otimes \tau_{0} \otimes \sigma_{0} \otimes \tau_{0}\right)\nonumber \\
&-&x\; x'\left(\sigma_{x} \otimes \tau_{0} \otimes \sigma_{x} \otimes \tau_{0}+\sigma_{y} \otimes \tau_{0} \otimes \sigma_{y} \otimes \tau_{0}\right) \nonumber \\
&-&w_{st}^2\left(\sigma_{x} \otimes \tau_{z} \otimes \sigma_{x} \otimes \tau_{z}+\sigma_{y} \otimes \tau_{z} \otimes \sigma_{y} \otimes \tau_{z}\right) \nonumber \\
&-&w_{ac}^{2}\;(\sigma_{z} \otimes \tau_{y} \otimes \sigma_{z} \otimes \tau_{y})\nonumber\\
& -&(m^{2}+w_{zz}^2)\;(\sigma_{z} \otimes \tau_{z} \otimes \sigma_{z} \otimes \tau_{z}), \label{DCr}
\end{eqnarray}
where ${\cal N}=N_T-2\pi i (\epsilon-\epsilon')+x^2+x'^2 +2 w_{st}^2+w_{ac}^2+m^2+w_{zz}^2$. Taking the inverse in Eq. (\ref{DCr}) we can calculate the average of the conductance from Eq.(\ref{Gm}), with ${\cal T}=\sigma_{0} \otimes \tau_{0} \otimes\sigma_{y} \otimes \tau_{0} $, we find the following general expression

\begin{eqnarray}
   \frac{ \left<G\right>}{e^2/h}&=& 4 \times \frac{N_1N_2}{N_T}-2\times\frac{N_1N_2}{N_T}\times\left[\frac{1}{N_{{\cal C}}+2w_{st}^2} \right.\nonumber\\
&+& \frac{1}{N_{ {\cal C}}+2w_{st}^2+2m^2+2w_{zz}^2}\nonumber\\
&+&\frac{1}{N_{{\cal C}}+2w_{st}^2+2w_{ac}^2+2m^2+2w_{zz}^2}\nonumber\\
&-&\left. \frac{1}{N_{{\cal C}}+2w_{st}^2+2w_{ac}^2}\right]
 \label{formulagwl}
\end{eqnarray}
where $N_{\cal C}=N_T+2x^2$ with $\epsilon = \epsilon'$ and $x=x'$. Eq. (\ref{formulagwl}) is the first major result of our work. The first term expresses Ohm's Law, while the remaining ones are known as the weak localization part of the average, $\left<G_{wl}\right>$.

\begin{figure}[!]
\begin{center}
\includegraphics[width=8cm,height=10cm]{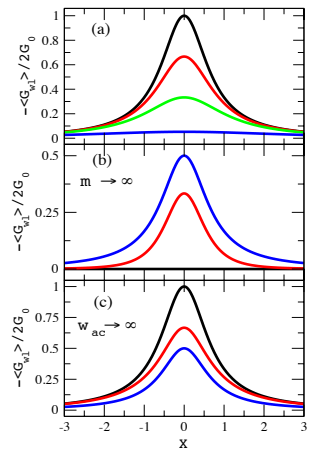}
\end{center}
\caption{(a) Eq.(\ref{formulagwlb}) is plotted as a function of the magnetic flux for ${\cal W}_{st} = 0, 0.5, 1, 3$ (from top to bottom). (b)  Eq. (\ref{formulagwlm}) is plotted as a function of the magnetic flux for  ${\cal W}_{ac} = 0, 1, \infty$ (from bottom to top). (c) Eq. (\ref{formulagwlac}) is plotted  as a function of the magnetic flux for  ${\cal M}= 0, 1, \infty$ (from top to bottom) and ${\cal W}_{zz}={\cal W}_{st}=0$.} \label{gwl}
\end{figure}

Let us analyze some relevant limits of Eq. (\ref{formulagwl}). As expected, the limit $x \to \infty $ leads to $\left<G_{wl}\right> \to 0$. A simple expression can be obtained by taking $m=w_{zz}=0$ in Eq. (\ref{formulagwl}):
\begin{eqnarray}
  \frac{ \left<G_{wl}\right>}{G_{0}}&=&
-\frac{2}{1+2{\cal X}^2+2{\cal W}_{st}^2},
 \label{formulagwlb}
\end{eqnarray}
which was obtained through a change of variables, $x^2={\cal X}^2N_T$, $b^2={\cal W}_{st}^2N_T$ and $G_0=e^2/h\times 2 \times N_1N_2/N_T^2$. From Eq. (\ref{formulagwlb}), we can conclude that weak localization is not affected by armchair edge ($w_{ac}$) if there is nomassive or zigzag edges present. In the Fig. (\ref{gwl}-a) we show Eq. (\ref{formulagwlb}) as a function of magnetic flux (${\cal X}$) for the following values of ${\cal W}_{st} = 0, 0.5, 1, 3$ (from top to bottom). Without straining (${\cal W}_{st}=0$), Fig. (\ref{gwl}-a) shows a weak localization peak. However, the peak becomes prominent with the increase in straining in monolayer CDQD. This result is in complete agreement with theoretical predictions of Ref. [\onlinecite{Guinea}] and with the experimental measurement of Ref. [\onlinecite{Geim1}],  which showed absence of a weak localization peak in monolayer of graphene because of straining (see Fig. ($2$-a) of Ref.[\onlinecite{Geim1}]).

In order to recover the results of Ref.[\onlinecite{Adagideli1}], we take the limit $m \to \infty $ in Eq. (\ref{formulagwl}). We obtain,
\begin{eqnarray}
   \frac{ \left<G_{wl}\right>}{G_0}&=&
-\frac{1}{1+2{\cal X}^2+2{\cal W}_{st}^2}\nonumber\\&+&\frac{1}{1+2{\cal X}^2+2{\cal W}_{st}^2+2{\cal W}_{ac}^2},
 \label{formulagwlm}
\end{eqnarray}
where $w_{ac}^2={\cal W}_{ac}^2N_T$. In the case without  straining (${\cal W}_{st}=0$), Eq. (\ref{formulagwlm}) reduces to Eq. (71) of Ref. [\onlinecite{Adagideli1}]. In Fig. (\ref{gwl}-b) we show Eq. (\ref{formulagwlm}) as a function of the magnetic flux (${\cal X}$) through the system for  values of ${\cal W}_{ac} = 0, 1, \infty$ (from bottom to top) and ${\cal W}_{st}=0$. For ${\cal W}_{ac} = 0$ the weak localization peak is absent, while it reaches a maximum  for ${\cal W}_{ac} \to\infty$.

In the limit $w_{zz} \to \infty $, Eq.(\ref{formulagwl}) reduces to Eq. (\ref{formulagwlm}). This is interesting, as massive and zigzag edges are physically different but serve the same purpose as far as weak localization is concerned, see Eq. (\ref{formulagwl}). This fact contributes to our conclusion that the armchair edge is only relevant in the presence of massive or zigzag edges.

The last important limit can be obtained from Eq. (\ref{formulagwl}) by letting $w_{ac} \to \infty $, 
\begin{eqnarray}
 \frac{ \left<G_{wl}\right>}{G_0}&=&
-\frac{1}{1+2{\cal X}^2+2{\cal W}_{st}^2}\nonumber\\&-&\frac{1}{1+2{\cal X}^2+2{\cal W}_{st}^2+2{\cal M}^2+2{\cal W}_{zz}^2},
 \label{formulagwlac}
\end{eqnarray}
where $m^2={\cal M}^2N_T$,$w_{zz}^2={\cal W}_{zz}^2N_T$. In Fig. (\ref{gwl}-c) we show Eq. (\ref{formulagwlac}) as a function of the magnetic flux (${\cal X}$) for ${\cal M}= 0, 1, \infty$ (from top to bottom) and ${\cal W}_{zz}={\cal W}_{st}=0$. For ${\cal M}\to \infty$ the weak localization peak decreases by a factor of two. The same conclusions are reached by fixing ${\cal M}={\cal W}_{st}=0$ and varying ${\cal W}_{zz}$.

\subsection{Covariance of Conductance}
Here we analyze how the weak localization peak is affected by the magnetic flux and edges. For this purpose we consider the covariance of conductance as a function of energy and magnetic flux using the same method described previously. From Eqs.(\ref{covg}) and (\ref{DCr}), we obtain the following general expression 
\begin{widetext}
\begin{eqnarray}
\frac{ {\bf cov}\left[G(\epsilon,x),G(\epsilon',x')\right]}{e^4/h^2}&=&4\times \frac{N_1^2N_2^2}{N_T^2} \times\left[\frac{1}{\left|N_{\cal D}\right|^2}+ \frac{1}{\left|N_{\cal D}+2w_{ac}^2\right|^2}+\frac{1}{\left|N_{\cal D}+4w_{st}^2+2m^2+2w_{zz}^2\right|^2}\right.\nonumber\\&+&\frac{1}{\left|N_{\cal D}+4w_{st}^2+2m^2+2w_{zz}^2+2w_{ac}^2\right|^2}\nonumber\\
&+&  \frac{1}{\left|N_{\cal C}+2w_{st}^2\right|^2}+\frac{1}{\left|N_{\cal C}+2w_{st}^2+2m^2+2w_{zz}^2\right|^2}+\frac{1}{\left|N_{\cal C}+2w_{st}^2+2w_{ac}^2\right|^2}\nonumber\\&+&\left.\frac{1}{\left|N_{\cal C}+2w_{st}^2+2m^2+2w_{zz}^2+2w_{ac}^2\right|^2}\right],\label{varianciaformula}
\end{eqnarray}
\end{widetext}
where $N_{{\cal D}}=N_T+2i\pi\left(\epsilon-\epsilon'\right)/\Delta+\left(x-x'\right)^2/2$ e $N_{{\cal C}}=N_T+2i\pi\left(\epsilon-\epsilon'\right)/\Delta+\left(x+x'\right)^2/2$. Eq. (\ref{varianciaformula}) is the second major result of our work. The first four terms of Eq. (\ref{varianciaformula}) are diffusons diagrams contributions that vanish in the presence of magnetic flux, while the remaining terms are cooperons diagrams contributions, which evanesce in the presence of magnetic flux  ($x\to\infty$).

\begin{figure}[!]
\begin{center}
\includegraphics[width=14cm,height=14cm]{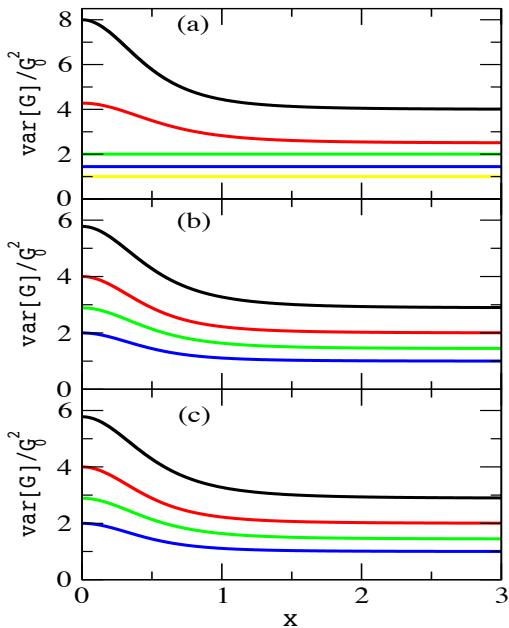}
\end{center}
\caption{(a) Eq. (\ref{varianciaformula}) is plotted (top to down) for ${\cal W}_{st}=0, 0.5,\infty$ (${\cal W}_{ac}={\cal W}_{zz}={\cal M}=0$) beyond Eq.  (\ref{varianciaformulaB}) for ${\cal W}_{ac}=0.5,\infty$. (b)  Eq. (\ref{varianciaformula}) is plotted (top to down)  for ${\cal M}=0.5,\infty$ (${\cal W}_{st}={\cal W}_{ac}={\cal W}_{zz}=0$) and Eq.  (\ref{varianciaformulam}) for ${\cal W}_{ac}=0.5,\infty$  (${\cal W}_{st}=0$). (c) Eq. (\ref{varianciaformula}) is plotted (top to down) for ${\cal W}_{ac}=0.5,\infty$ (${\cal W}_{st}={\cal M}={\cal W}_{zz}=0$), and Eq.  (\ref{varianciaformulaac}) for ${\cal M},{\cal W}_{zz}=0.5,\infty$  (${\cal W}_{st}=0$).} \label{varg}
\end{figure}

Next we analyze the same limits of  Eq. (\ref{varianciaformula}). Taking $\epsilon=\epsilon'$, $x=x'$, without magnetic flux and setting the others equal to zero, the variance of the conductance from Eq. (\ref{varianciaformula}) is given by ${\bf var}\left[G\right]=G_0^2 \times \left[ 4 \times 2 \right] $, where the factor $4$ is the degeneracy of the sub-lattice and valley symmetries and the factor $2$ comes about from time-reversal symmetry. Further, in the presence of magnetic flux, the variance of the conductance is given by ${\bf var}\left[G\right]=G_0^2 \times \left[ 4 \times 1 \right] $, which indicates that time-reversal symmetry is broken, see  top curve of Fig. (\ref{varg}-a).

Simple expression can been obtained taking $w_{st}\to\infty$ in Eq.(\ref{varianciaformula}):
\begin{eqnarray}
\frac{ {\bf var}\left[G\right]}{G_0^2}&=&\sum_{i=0}^1\frac{1}{\left(1+2i{\cal W}_{ac}^2\right)^2}\label{varianciaformulaB}.
\end{eqnarray}
Note that, two diffuson and all cooperon contributions  have vanished, indicating the breaking time-reverse symmetry. From Eq. (\ref{varianciaformulaB}), only armchair edges are relevant in the presence of straining.  Moreover, the average of variances are given by  ${\bf var}\left[G\right]=G_0^2 \times \left[ 2 \times 1 \right] $ and ${\bf var}\left[G\right]=G_0^2 \times \left[ 1 \times 1 \right] $ for ${\cal W}_{ac}=0$ and ${\cal W}_{ac}\to\infty$, respectively. In Fig. (\ref{varg}-a)  we show (top to down) Eq. (\ref{varianciaformula}) for ${\cal W}_{st}= 0, 0.5,\infty$ (${\cal W}_{ac}={\cal W}_{zz}={\cal M}=0$), and Eq.  (\ref{varianciaformulaB}) for ${\cal W}_{ac}=0.5,\infty$.

\begin{figure}[!]
\begin{center}
\includegraphics[width=8.5cm,height=6cm]{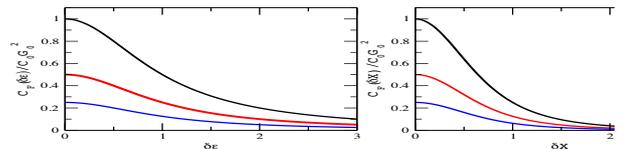}
\end{center}
\caption{Typical Lorentzian (left) and quadratic Lorentzian correlation function from Eq. (\ref{CF}) for: (top to down) all parameters set to zero,  $\{{\cal X} ,{\cal M},{\cal W}_{ac},{\cal W}_{zz} \to \infty\}$  and ${\cal W}_{st} \to \infty$.} \label{Lo}
\end{figure}

Taking $m\to\infty$ ( or $w_{zz}\to\infty$)  in Eq. (\ref{varianciaformula}), four terms go to zero, two diffusons and two cooperons contributions. In this case,  Eq.(\ref{varianciaformula}) simplifies to
\begin{eqnarray}
\frac{ {\bf var}\left[G\right]}{G_0^2}&=&\sum_{i,j=0}^1\frac{1}{\left(1+2i{\cal X}^2+2i{\cal W}_{st}^2+2j{\cal W}_{ac}^2\right)^2}\label{varianciaformulam}.
\end{eqnarray}
In the case without  straining (${\cal W}_{st}=0$),  Eq. (\ref{varianciaformulam}) reduces to Eq. (84) of Ref. [\onlinecite{Adagideli1}]. Without magnetic flux and setting the other parameters equal to zero, the variance of the conductance from Eq. (\ref{varianciaformulam}) is given by ${\bf var}\left[G\right]=G_0^2 \times \left[ 2 \times 2 \right] $,  which means that the degeneracy factor is reduced by a factor of two and time-reversal symmetry is not  broken by the massive edge. Moreover, with magnetic flux on, the variance of the conductance is given by ${\bf var}\left[G\right]=G_0^2 \times \left[ 2 \times 1 \right] $. On the other hand, if ${\cal W}_{ac}\to\infty$ and ${\cal W}_{st}=0$ the variance from Eq.(\ref{varianciaformulam}) goes to ${\bf var}\left[G\right]=G_0^2 \times \left[ 1 \times 2 \right] $ and ${\bf var}\left[G\right]=G_0^2 \times \left[ 1 \times 1 \right] $ without and with magnetic flux, respectively.  In Fig. (\ref{varg}-b)  we show (top to down) Eq.(\ref{varianciaformula}) for ${\cal M}=0.5,\infty$ (${\cal W}_{st}={\cal W}_{ac}={\cal W}_{zz}=0$) and Eq.(\ref{varianciaformulam}) for ${\cal W}_{ac}=0.5,\infty$  (${\cal W}_{st}=0$).

The last limit we consider, $w_{ac}\to\infty$, in Eq.(\ref{varianciaformula}), gives,
\begin{eqnarray}
\frac{ {\bf var}\left[G\right]}{G_0^2}&=&\sum_{i=0}^1\frac{1}{\left(1+4i{\cal W}_{st}^2+2i{\cal M}^2+2i{\cal W}_{zz}^2\right)^2}\nonumber\\
&+&\sum_{i=0}^1\frac{1}{\left(1+2{\cal X}^2+2{\cal W}_{st}^2+2i{\cal M}^2+2i{\cal W}_{zz}^2\right)^2}\label{varianciaformulaac}.
\end{eqnarray}
In this case, the contributions of two diffusons and two cooperons vanish. Fixing ${\cal W}_{st}=0$ and turning off the magnetic flux, the variance of Eq. (\ref{varianciaformulaac}) becomes ${\bf var}\left[G\right]=G_0^2 \times \left[ 2 \times 2 \right] $ and ${\bf var}\left[G\right]=G_0^2 \times \left[ 1 \times 2 \right] $ for ${\cal M},{\cal W}_{zz}=0$ and ${\cal M},{\cal W}_{ac}\to\infty$, respectively, which indicates that time-reversal symmetry is preserved in both cases. On the other hand, with magnetic flux turned on, the variance of Eq. (\ref{varianciaformulaac}) goes to  ${\bf var}\left[G\right]=G_0^2 \times \left[ 2 \times 1 \right] $ and ${\bf var}\left[G\right]=G_0^2 \times \left[ 1 \times 1 \right] $ for ${\cal M},{\cal W}_{zz}=0$ and ${\cal M},{\cal W}_{ac}\to\infty$, respectively, indicating the breaking of time-reverse symmetry in both cases.  In Fig. (\ref{varg}-c)  we show (top to down) Eq. (\ref{varianciaformula}) for ${\cal W}_{ac}=0.5,\infty$(${\cal W}_{st}={\cal M}={\cal W}_{zz}=0$) and Eq.(\ref{varianciaformulaac}) for ${\cal M},{\cal W}_{zz}=0.5,\infty$  (${\cal W}_{st}=0$).

\subsection{Correlation Function}

After analyzing in detail the variance of the conductance of the chaotic Dirac quantum dot from Eq. (\ref{varianciaformula}), we briefly study how the correlation function ${\cal C_F}(\delta \epsilon,\delta {\cal X})$, or covariance of conduction,  is affected by straining and boundary parameters. Substituting $\epsilon'=\epsilon+\delta\epsilon$ and $x'=x+\delta x$ in  Eq. (\ref{varianciaformula}), we can write
\begin{eqnarray}
\frac{{\cal C_F}(\delta \epsilon,\delta {\cal X})}{G_0^2}&=&C_\lambda \times \frac{1}{\left|1+i\delta\epsilon+\delta {\cal X}^2\right|^2},\label{CF}
\end{eqnarray}
where $C_\lambda $ is a constant ($\lambda=\{0,{\cal X},{\cal W}_{st},{\cal M},{\cal W}_{ac},{\cal W}_{zz}\}$), while $C_ 0= 4 \times 2$, with all parameters set to zero, $C_ {\cal X}= 4 \times 1$ with $ {\cal X} \to \infty$ and all other parameters set to zero, $C_{\lambda}=2 \times 2$ with ${\cal M},{\cal W}_{ac},{\cal W}_{zz} \to \infty$ and the other parameters being zero, $C_{{\cal W}_{st}}=2 \times 1$ with ${\cal W}_{st} \to \infty$ and other parameters are set to zero. For $\delta  {\cal X}=0$, the correlation function is a typical Lorentzian:
$$ \frac{{\cal C_F}(\delta \epsilon)}{G_0^2} = C_\lambda \times  \frac{1}{1+\delta\epsilon^2},$$
which is in accord with the experiment of Ref. [\onlinecite{Lerner}]. Moreover, for $\delta \epsilon=0$ the correlation function is a quadratic Lorentzian
$$ \frac{{\cal C_F}(\delta{\cal X})}{G_0^2} = C_\lambda \times   \frac{1}{\left(1+\delta {\cal X}^2\right)^{2}},$$
which is in agreement with the result of analysis in the experiment of Ref. [\onlinecite{ Folk}]. Lorentzian and quadratic Lorentzian shapes of the correlation function are  plotted in Fig. (\ref{Lo}). These findings are encouraging as they confirm the premise of this paper that Chaotic Dirac Quantum Dots containing relativistic electrons obeying the Dirac equation, exhibit universal fluctuations describable by RMT.

\section{Conclusions}
\label{Con}

In this paper the random nature of the conductance in chaotic Dirac quantum dots is investigated using the RMT-based stub mode. Analytical results for the average conductance and the correlation function are obtained and scrutinized under different limiting  situations. The results coincide with those obtained using the semiclassical approach and, when available, agree with experimental findings. Accordingly, the chaotic graphene quantum dot, also called the chaotic Dirac quantum dot, with the electrons motion governed by the Dirac equation  is a mesoscopic system that follows the rules of RMT, just as the chaotic Schr\"odinger quantum dot.

\begin{acknowledgments}
This work was partially supported by the nBrazilian agencies CNPq, FAPESP, CAPES, INCT-IQ/MCT, FACEPE and CEPID-CEPOF\end{acknowledgments}

\end{document}